\documentclass[a4paper,11pt]{article}
\usepackage{jheppub,xcolor} 

\usepackage[T1]{fontenc} 

\newcommand{\cR}{{\mathcal{R}}}

\newcommand{\nn}{\nonumber}

\def \cf{{\cal F}}

\title{\rm \bf \Huge Ghostbusters in $f(R)$ supergravity}

\author[a]{Toshiaki Fujimori,}
\author[a]{Muneto Nitta,}
\author[a]{Keisuke Ohashi,}
\author[b]{and Yusuke Yamada}
\affiliation[a]{Department of Physics, \& Research and Education Center for Natural Sciences, Keio University, Hiyoshi 4-1-1, Yokohama, Kanagawa 223-8521, Japan}
\affiliation[b]{Stanford Institute for Theoretical Physics and Department of Physics, Stanford University, Stanford, CA 94305, USA}

\abstract{
$f(R)$ supergravity is known to contain a ghost mode 
associated with higher-derivative terms 
if it contains $R^n$ with $n$ greater than two.
We remove the ghost 
 in $f(R)$ supergravity 
 by introducing auxiliary gauge field 
 to absorb the ghost.
 We dub this method as the ghostbuster mechanism
\cite{Fujimori:2016udq}.
We show that the mechanism removes the ghost supermultiplet but also terms including $R^n$ with $n\geq3$, after integrating out auxiliary degrees of freedom. For pure supergravity case, there appears an instability in the resultant scalar potential. 
We then show that the instability of the scalar potential 
can be cured 
by introducing matter couplings in such a way that 
the system has a stable potential.}

\begin{document} 
\maketitle
\flushbottom

\section{Introduction}
Higher-order derivative interactions naturally appear 
in effective field theories. 
In particular, in the system with gravity, 
we need to take into account such terms 
since various higher-order corrections can be 
relevant to the dynamics. 
However, higher-derivative interactions often lead to 
the so-called Ostrogradski instability 
\cite{Ostrogradski,Woodard:2006nt}: 
higher-derivative interactions give 
additional degrees of freedom 
which makes the Hamiltonian unbounded from below, 
and hence the system shows an instability. 
If such a ghost mode appears, 
one should regard the system as an effective theory 
which is valid only below 
the energy scale of the mass of the ghost mode, 
otherwise the system loses the unitarity. 
In a class of ghost-free higher-derivative interactions, 
one does not come across with such an instability problem. 
In the case of a system with a single scalar and a tensor, 
the Horndeski class
~\cite{Horndeski:1974wa,Kobayashi:2011nu} of interactions 
are free from ghosts. 
In this class of interactions, 
the equations of motion (E.O.M) are 
at most the second order differential equations, 
and no additional degree of freedom shows up. 
In general,
one may ask the following question: 
among many possible higher-order derivative terms, 
what kind of structure gives us ghost-free interactions? 
For example, in the so-called Galileon models~\cite{Nicolis:2008in}, Galileon scalar fields can be understood 
as the Goldstone mode of translation symmetry in extra dimensions, and the action is made out of ghost-free derivative terms. 
Therefore, one can say that the hidden translation symmetry 
controls the higher-derivative interactions 
so that there appear no new degrees of freedom.  
The absence of ghosts  
in supersymmetric Galileon model~\cite{Farakos:2013zya} 
can be also achieved 
by a spontaneously broken hidden SUSY~\cite{Roest:2017uga}.

Higher-derivative interactions are also studied in gravity theories. 
Despite the existence of fourth-order derivative interactions, 
the so-called Starobinsky model~\cite{Starobinsky:1980te}, 
which has a quadratic term of the Ricci scalar, 
does not have any ghost as well as the Horndeski class. 
This is because such a system is equivalent to 
the scalar-tensor system without higher-derivatives. As a cosmological application,
 the Starobinsky model predicts 
the spectral tilt of scalar curvature perturbation compatible with 
the latest CMB observation~\cite{Ade:2015lrj}. 
One can extend this model to the system with 
an arbitrary function of the Ricci scalar, called the $f(R)$-gravity model~\cite{Buchdahl:1983zz} (see also Ref.~\cite{DeFelice:2010aj,Nojiri:2017ncd} for review),
which is also dual to a scalar-tensor system, 
and therefore free from the ghost instability.

Higher-derivative interactions were also studied 
in supersymmetric (SUSY) theories, 
both for global SUSY and supergravity (SUGRA).
In SUSY cases, there is another 
problem called the auxiliary field problem:
space-time derivatives may act in general on 
SUSY auxiliary fields 
($F$ and $D$ for chiral and vector multiplets, respectively) 
in the off-shell superfield formulation.
Then, they become dynamical and so 
one cannot eliminate them by their E.O.M 
\cite{Gates:1995fx-0,Gates:1995fx-1}. 
The auxiliary field problem and the higher-derivative ghosts usually 
come up together~\cite{Antoniadis:2007xc,Dudas:2015vka}.
In four dimensional (4D) ${\cal N}=1$ SUSY theories, 
a classification of 
higher-derivative terms free from ghosts 
and the auxiliary field problem 
was given for chiral superfields 
\cite{Khoury:2010gb,Khoury:2011da,Koehn:2012ar,Koehn:2012te} 
as well as for vector superfields \cite{Fujimori:2017kyi}. 
Such higher-derivative interactions of chiral superfields 
were applied to 
low-energy effective theory~\cite{Buchbinder:1994iw-0,Buchbinder:1994iw-1,Buchbinder:1994iw-2,Buchbinder:1994iw-3} 
(see also \cite{Nitta:2014fca}),  
coupling to SUGRA
\cite{Koehn:2012ar,Farakos:2012qu},   
 Galileons~\cite{Khoury:2011da}, 
 ghost condensation \cite{Koehn:2012te}, 
a Dirac-Born-Infeld (DBI) inflation~\cite{Sasaki:2012ka}, 
flattening of the inflaton potential~\cite{Aoki:2014pna-0,Aoki:2014pna-1},
a (baby) Skyrme model 
\cite{Adam:2013awa-0,Adam:2013awa-1,Nitta:2014pwa-0,Nitta:2014pwa-1,Bolognesi:2014ova-0,Bolognesi:2014ova-1, 
Gudnason:2015ryh-0,Gudnason:2015ryh-1,Queiruga:2015xka},
other BPS solitons \cite{Nitta:2014pwa-0,Nitta:2014pwa-1,Queiruga:2017blc,Eto:2012qda}, 
and modulated vacua \cite{Nitta:2017mgk,Nitta:2017yuf-1},
while 
higher-derivative interactions of vector superfields  
were applied to 
 the DBI action  
\cite{Cecotti:1986gb,Bagger:1996wp,Rocek:1997hi}, 
SUGRA coupling 
\cite{Cecotti:1986gb,Kuzenko:2002vk-0,Kuzenko:2002vk-1,Abe:2015nxa},
SUSY Euler-Heisenberg action 
\cite{Farakos:2012qu,Cecotti:1986jy,Farakos:2013zsa,Dudas:2015vka}, 
and non-linear self-dual actions 
\cite{Kuzenko:2002vk-0,Kuzenko:2002vk-1,Kuzenko:2000tg-0,Kuzenko:2000tg-1,Kuzenko:2000tg-2}. 
 
On the other hand, higher-derivative interaction of gravity multiplets 
were studied in 4D ${\cal N}=1$ SUGRA. In Ref.~\cite{Cecotti:1987sa},  
Cecotti constructed the higher-order terms of the Ricci scalar in the old minimal supergravity formulation
and showed that
at least one ghost superfield appears  
if we have $R^n$ $(n\geq3)$ terms in the system.
It is possible to avoid the ghost 
by some modifications of the system. 
In \cite{Ferrara:2014fqa}, 
the so-called nilpotent constraint 
on the Ricci scalar multiplet, 
which removes a scalar field in the multiplet, 
is considered. 
Due to the absence of the scalar, 
the bosonic ghost is absent 
in the spectrum of the system. 
This mechanism has been applied 
to various higher-curvature models 
in SUGRA~\cite{Farakos:2017mwd}. 
The nilpotent constraint $\cR^2=0$, 
however, is an effective description of a broken-SUSY system. 
If the linearly realized SUSY is restored in a higher energy regime, 
the ghost mode would show up.\footnote{
The nilpotent condition on a chiral superfield $\Phi$ has two solutions. 
A nontrivial solution is $\phi=\frac{\psi\psi}{F^\phi}$ 
where $\phi$, $\psi$ and $F^\phi$ are scalar, Weyl spinor, and auxiliary scalar components of $\Phi$. 
Obviously, this solution is well-defined for $F^\phi\neq0$, 
that is, SUSY should be spontaneously broken.  
}
As another approach, in~\cite{Diamandis:2017ems} 
the authors considered a deformation of the ghost kinetic term 
by introducing an additional K\"ahler potential term. 
It is shown that the resultant ghost-free system is equivalent 
to the matter coupled $f(R)$ SUGRA.

Meanwhile,
in our previous work~\cite{Fujimori:2016udq}, 
we proposed a simple method to remove a ghost mode in 
4D ${\cal N}=1$ SUSY chiral multiplets 
\cite{Antoniadis:2007xc,Dudas:2015vka}, 
which we dubbed ``ghostbuster mechanism.'' 
We gauge a U(1) symmetry 
by introducing a non-dynamical gauge superfield without kinetic term 
to the higher-derivative system 
with assigning charges on chiral superfields properly 
in order for  
the gauge field to absorb the ghost.
Namely, due to the gauge degree of freedom,
the ghost in the system is removed 
by the U(1) gauge fixing.
In this class of models, 
a hidden local symmetry plays a key role 
in the ghostbuster mechanism.
Actually, before this work, 
esentially the same technique is used 
for superconformal symmetry 
in the conformal SUGRA formalism: 
the conformal SUGRA has one ghost-like degree of freedom 
called as a compensator. 
Such a degree of freedom is removed 
by the superconformal gauge fixing, 
whereas in the ghostbuster mechanism, 
the hidden local U(1) gauge fixing removes 
the ghost associated with higher-derivatives. 
Therefore, in SUGRA models, 
one may understand the higher-derivative ghost 
as a second compensator for the system 
with the superconformal symmetry 
$\times$ hidden local U(1) symmetry.

In this paper, we apply the ghostbuster mechanism 
to remove the ghost in the $f(R)$ SUGRA system. 
Interestingly, the hidden U(1) symmetry 
required for the mechanism 
can be understood as the gauged R-symmetry, 
since the gravitational superfield 
should be gauged under the U(1) symmetry. 
The U(1) charge assignment is uniquely determined, 
and therefore, naively one cannot expect  
a ghost mode cancelation a priori. 
As we will show, 
a would-be ghost superfield 
has a gauge charge and can be nicely removed 
by the gauge fixing of the U(1) symmetry.
As a price of this achievement, however, 
the resultant system generically has 
an unstable scalar potential in a pure SUGRA case. 
Such an unstable scalar potential 
can be cured by various modifications. 
As an example we propose 
a model with a matter chiral superfield. 
We will find that such a deformation 
leads to a healthy model of SUGRA 
without either ghosts or instabilities of the scalar potential.

One will easily find how the ghost supermultiplet is eliminated from the dual matter-coupled SUGRA viewpoint. We also address the same question in the higher-curvature SUGRA system. We find that, after integrating out the auxiliary vector superfield for the mechanism, the scalar curvature terms including $R^n$ with $n\geq3$ disappear, and the resultant system has linear and quadratic terms in $R$. However, the $R+R^2$ SUGRA system has couplings completely different from that proposed in \cite{Cecotti:1987sa}. This observation means that, despite the disappearance of higher scalar curvatures in the final form, the higher-curvature deformation in the original action gives a physical consequence even after applying the ghostbuster mechanism.

This paper is organized as follows. 
In Sec.\,\ref{review}, we briefly review 
the higher-curvature SUGRA models 
and its dual description. 
In particular, one finds that 
once the SUSY version of 
the higher order Ricci scalar term $R^n$ ($n\geq3$) is included in the old minimal SUGRA formulation, 
there appears at least one ghost chiral superfield. 
We apply the ghostbuster mechanism 
to the higher-curvature SUGRA in Sec.\,\ref{GB}. 
We will see that although the ghost superfield 
can be removed by the mechanism, 
the resultant system has a scalar potential 
with an instability in the direction of a scalar field. 
Then, in Sec.\,\ref{sec:unstablemodel}, 
we discuss a simple modification of the model 
by introducing an extra matter chiral superfield. 
We show an example which is stable and free from ghost as well. 
Finally, we conclude in Sec.\,\ref{conclusion}. 
Throughout this paper, 
we will use the notation of~\cite{Freedman:2012zz}.


\section{Higher-curvature terms in supergravity}
\label{review}
In this section, we review 
the construction of higher-order terms of the Ricci scalar 
in 4D ${\cal N}=1$ SUGRA~\cite{Cecotti:1987sa}.\footnote{Cosmological application of SUSY Starobinsky model is discussed e.g. in~\cite{Kallosh:2013xya,Farakos:2013cqa,Ketov:2013dfa}.}
In this paper, we use the conformal SUGRA formalism, 
in which there are conformal symmetry and 
its SUSY counterparts 
in addition to super-Poincar\'e symmetry 
\cite{Kaku:1978nz,Kaku:1978ea,Townsend:1979ki,Kugo:1982cu}. 
In order to fix the extra gauge degree of freedom, 
we need to introduce an unphysical degrees of freedom 
called the conformal compensator, 
which should be in a superconformal multiplet. 
In this paper, we adopt a chiral superfield 
as a compensator superfield, 
which leads to the so-called old minimal SUGRA 
after superconformal gauge fixing. 
We show the components of supermultiplet, 
the density formulas, and identities in Appendix~\ref{Components}.

First, let us show the pure conformal SUGRA action,
\begin{align}
S = \left[ - \frac{3}{2} S_0 \bar{S}_0 \right]_D, \label{pure}
\end{align}
where $S_0$ is the chiral compensator with the charges $(w,n)=(1,1)$ in 
conformal SUGRA (see Apeendix \ref{Components} for the definition of the charges), 
and $[\cdots]_D$ denotes the D-term density formula. 
Taking the pure SUGRA gauge, 
$S_0=\bar{S}_0=1$, $b_\mu=0$, 
we obtain an action whose bosonic part takes the form
\begin{align}
S=\int d^4x\sqrt{-g}\left(\frac{1}{2}R-3|F^{S_0}|^2+3A_aA^a\right),
\end{align}
where $R$ is the Ricci scalar, 
$F^{S_0}$ is the F-term of $S_0$ 
and $A_a$ is the gauge field of chiral U(1)$_A$ symmetry, 
which is a part of superconformal symmetry. 
The E.O.M.  
for the auxiliary fields $F^{S_0}$ and $A_a$ 
can be solved by setting $F^{S_0} = A_a = 0$, 
and then we find the pure SUGRA action. 
The action~(\ref{pure}) can also be written as
\begin{align}
S = \left[ \frac{3}{2} S_0^2 \cR \right]_F,
\end{align}
where $[\cdots]_F$ is the F-term density formula. 
Here we have used the identity given in~\eqref{ID}. 
The chiral superfield $\cR$ is 
the so-called scalar curvature superfield, 
defined by
\begin{align}
\cR \equiv \frac{\Sigma(\bar{S}_0)}{S_0},
\end{align}
where $\Sigma$ is the chiral projection operator. 
Its components in the pure SUGRA gauge are given by
\begin{align}
\cR = [ \Phi \,,\, P_L \chi \,,\, F ] 
= \left[ -\bar{F}^{S_0} \,,\, \cdots \,,\, 
|F^{S_0}|^2 + \frac{1}{6} R + A_a A^a - i \partial_a A^a 
+ \cdots \right],
\end{align}
where ellipses denote fermionic parts. 
From this expression, 
we find that the F-component of $\cR$ contains the Ricci scalar.

It has been known that there is no ghost 
in the system involving $R^2$, which is realized as
\begin{align}
S = \left[ - \frac{3}{2} S_0 \bar{S}_0 
+ \frac{\alpha}{2} \cR \bar{\cR} \right]_D, \label{Rsquare}
\end{align} 
where $\alpha$ is a real constant. 
The bosonic part of this action 
after the superconformal gauge fixing is
\begin{align}
S|_B
= \int d^4x \sqrt{-g} \Biggl[&
\frac{R}{2} + \frac{\alpha}{36} R^2 
- 3 |F^{S_0}|^2 - \alpha D_a F^{S_0} D^a \bar{F}^{S_0}
+ 3 A_a A^a + \alpha (\partial_a A^a)^2 \nonumber \\
& + \frac{\alpha R}{6} \left( |F^{S_0}|^2 + 2 A_a A^a \right)
+ \alpha \left( |F^{S_0}|^2 + A_a A^a \right)^2 \Biggr],
\end{align}
where $D_a$ represents the covariant derivative, 
$D_a S_0 = ( \partial_a - iA_a ) S_0 = - i A_a, 
D_a F^{S_0} = ( \partial_a + 2i A_a ) F^{S_0}$.
The Lagrangian has the quadratic Ricci scalar term 
$\frac{\alpha}{36} R^2$ 
and also the non-minimal couplings 
between $F^{S_0}, A_a$ and $R$.  
In this system, there exist four real massive modes $\varphi_i$ 
with the common mass $m^2=3/\alpha$
in the fluctuations around the vacuum 
$g_{\mu\nu}=\eta_{\mu\nu} $ and $F^{S_0}=A_a=0$: 
\begin{eqnarray}
g_{\mu\nu}=\eta_{\mu\nu}+\left(\eta_{\mu\nu} -\frac{\partial_\mu\partial_\nu}{\Box}\right) \varphi_1, 
\quad    A_\mu =\partial_\mu \varphi_2, \quad  F^{S_0}=\varphi_3+i \varphi_4.
\end{eqnarray}
We stress that, 
as is often the case with SUSY higher derivative models, 
the auxiliary fields have their kinetic terms 
and hence they are dynamical degrees of freedom
in the presence of the higher-derivative term.

Next, let us consider a SUGRA system 
with $R^n$, $n\geq3$ along the line of 
Refs.~\cite{Cecotti:1987sa,Ozkan:2014cua,Ferrara:2014fqa}. 
As we discussed in the previous section, 
$\cR$ superfield has the Ricci scalar in its F-component. 
Using the chiral projection operator $\Sigma$, 
one can obtain the superfield $\Sigma(\bar{\cR})$ 
which has $R$ in the lowest component: 
\begin{align}
\Sigma(\bar{\cR}) =& \left[
- \frac{1}{6} R - |F^{S_0}|^2 - A_a A^a - i\partial_a A^a + \cdots \,,\,
\cdots \,,\, \right. \nonumber \\
& \hspace{1cm} \left. 
\frac{1}{6} R F^{S_0} + 
\left( \partial_a^2 + i \partial_a A^a - A_a A^a \right) F^{S_0} 
+ \cdots \right],
\end{align}
where we have shown only the relevant part. 
With this superfield $\Sigma(\bar{\cR})$, 
one can construct an action
involving arbitrary functions of $R$, 
i.e.~$f(R)$ gravity models in SUGRA. 
Here we consider the action of the form
\begin{align}
S=\left[-\frac{3}{2}S_0\bar{S}_0\Omega\left(\frac{\cR}{S_0},\frac{\bar{\cR}}{\bar{S}_0},\frac{\Sigma(\bar{\cR})}{S_0^2}, \frac{\bar{\Sigma}(\cR)}{\bar{S}_0^2}\right)\right]_D+\left[S_0^3\cf\left(\frac{\cR}{S_0},\frac{\Sigma(\bar{\cR})}{S_0^2}\right)\right]_F,\label{fR1}
\end{align}
where $\Omega$ is an arbitrary real function and 
$\cf$ is an arbitrary holomorphic function. 
If we chose $\Omega = 0, \cf(S,X) = S ( 3 - \alpha X)/2$,
then this action reduces to (\ref{Rsquare}) since
\begin{eqnarray}
\left[S_0^3\cf\left(\frac{\cR}{S_0},\frac{\Sigma(\bar{\cR})}{S_0^2}\right)\right]_F=
\left[\frac32 S_0^2 \cR-\frac \alpha 2 \cR \Sigma(\bar {\cal R})\right]_F=\left[-\frac32 S_0\bar S_0+\frac\alpha 2 \cR \bar \cR \right]_D.
\end{eqnarray}
The bosonic part of the action contains
the following terms including higher-order terms of Ricci scalar $R$
\begin{eqnarray}
\int d^4x \sqrt{-g} 
\left\{ - \frac{R^2}{12} \Omega_{S\bar S}(S,\bar S,X,\bar X) 
+ \frac{R}{6} \cf_S(S, X) + {\rm h.c.} 
\right\}_{S = -\bar F^{S_0} ,\, X= -R/6} \, , \label{curvature}
\end{eqnarray}
where the subscripts on the functions 
denote the differentiations with respect to the scalar fields.

Such SUSY higher-derivative terms have derivative interactions of auxiliary fields, and the interactions make the auxiliary fields dynamical as
\begin{eqnarray}
 \int d^4x \sqrt{-g}\left\{  
\frac{1}{12} g^{\mu\nu} 
\partial_\mu R \, \partial_\nu R \, \Omega_{X\bar X} +
\left( \partial_\mu^2 F^{S_0} \cf_{X} + {\rm h.c.} \right) + \cdots \right\}_{S = -\bar F^{S_0} ,\, X=-R/6} \, . 
\end{eqnarray}
In this system, 
in addition to the scalar degree of freedom 
from the derivative terms of the Ricci-curvature, 
the higher-derivative terms of 
the ``dynamical'' auxiliary field $F^{S_0}$ 
give rise to multiple scalar degrees of freedom, 
some of which are ghost-like.
If we choose
$\Omega(S,\bar S,X,\bar X) = S \bar S \tilde \Omega(X,\bar X)$, 
$\cf(S,X) = S \tilde \cf(X)$, 
and set $F^{S_0}=0$ identically 
as is done by imposing the nilpotent condition $\cR^2=0$ 
in Ref.~\cite{Ferrara:2014fqa}, 
the above terms vanish and no ghost seems to appear.  
Without such a condition, however, 
the appearance of ghost is unavoidable 
as is clearly shown in the following.

The present system is also equivalent to 
a standard SUGRA model coupled to matter superfields. 
As in the previous section, 
we use Lagrange multiplier suerfields, 
and rewrite the action~(\ref{fR1}) as
\begin{align}
S' = & 
\Bigg[
- \frac{3}{2} S_0 \bar{S}_0 \, \Omega(S,\bar{S},X,\bar{X}) 
\Bigg]_D 
+ \Bigg[ S_0^3 \, \cf(S,X) \Bigg]_F \nn\\
& \hspace{1cm}
+ \Bigg[ 3 S_0^3 \, T \left( \frac{\cR}{S_0} - S \right) \Bigg]_F 
+\Bigg[ 3 S_0^3 \,Y \left( \frac{\Sigma( \bar{S}_0 \bar{S} )}{S_0^2} 
- X \right) \Bigg]_F, 
\end{align}
where $T$ and $Y$ are Lagrange multiplier superfields with $(w,n)=(0,0)$. The E.O.Ms of $T$ and $Y$ give 
the constraints which reproduce the original action~(\ref{fR1}). 
Instead, using the identity~(\ref{ID}), 
we can also obtain the dual action
\begin{align}
S' = \left[ - \frac{3}{2} S_0 \bar{S}_0 
\left( T + \bar{T} + Y \bar{S} + \bar{Y} S 
+ \Omega(S,\bar{S},X,\bar{X}) \right) \right]_D
+ \bigg[S_0^3 \left( \cf(S,X) - 3TS - 3XY \right) \bigg]_F.\label{fR2}
\end{align}
This is a standard SUGRA system 
with the following K\"ahler and super-potentials,
\begin{align}
K &= -3 \log \left( T + \bar{T} + Y \bar{S} + \bar{Y} S + 
\Omega(S,\bar{S},X,\bar{X}) \right),
\label{eq:GhostKahler}\\
W &= \cf(S,X) - 3TS - 3XY.
\end{align}

Let us show the existence of a ghost mode. 
The K\"ahler metric of the $\{S,Y\}$ sector takes the form,
\begin{align}
K_{I\bar{J}}=\left(
\begin{array}{cc}
K_{S\bar{S}}&-\frac{1}{A}\\
-\frac{1}{A}&0
\end{array}
\right),
\end{align}
where $A=T+\bar{T}+Y\bar{S}+\bar{Y}S+\Omega(S,\bar{S},X,\bar{X})$. The determinant of this sub matrix has negative determinant, and 
this K\"ahler metric has one negative eigenvalue 
corresponding to a ghost. 
Thus, the $f(R)$ SUGRA model has one ghost mode in general.

Note that $X$ becomes an auxiliary superfield 
if $\Omega=\Omega(S,\bar{S})$ is independent of $X$. 
Even in such a case, 
the system has higher-curvature terms 
in the $\cf(S,X)$ term in \eqref{curvature}. 
The reduced dual system is described by
\begin{align}
K &= -3 \log 
\left( T+\bar{T}+Y\bar{S}+\bar{Y}S+\Omega(S,\bar{S}) \right), \nn \\
W &= g(S,Y) - 3TS, \label{redfr}
\end{align}
where $g(S,Y) = [\cf - X \cf_X]_{X=X(S,Y)}$ and 
$X(S,Y)$ is a solution of $\cf_X - 3Y = 0$.\footnote{
Here we assume that the equation $\cf_X = \cf_X(X)$ 
can be solved for $X$ (e.g. $\cf \propto  S X^{n-1}$ with $n\ge 3$). 
Constant and linear terms in $\cf$ merely rescale the $R$ and $R^2$ terms respectively.} This reduction does not change the above discussion, and hence a ghost mode appears in this system as well.
\section{Ghostbuster in $f(R)$ supergravity}
\label{GB}
In this section, 
we consider the elimination of the ghost superfield 
along the line of Ref.~\cite{Fujimori:2016udq}. 
To eliminate the ghost superfield, 
one needs to introduce a gauge redundancy, 
by which one of the degrees of freedom is removed. 
In the $f(R)$ SUGRA discussed above, 
all the superfields $\cR$, $\Sigma(\bar{\cR})$ are expressed 
in terms of $S_0$ with the SUSY derivative operators. 
Hence, once we introduce a vector superfield $V_R$ 
for a U(1) gauge symmetry 
and assign the charge to $S_0$ so that it transforms as
\begin{eqnarray}
S_0 \to e^{\Lambda} S_0, ~~~~~
V_R \to V_R - \Lambda - \bar\Lambda,
\end{eqnarray} 
the transformation law of $\cR$ and $\Sigma(\bar{\cR})$ 
are automatically determined as
\begin{eqnarray} 
\cR_g \equiv \frac{\Sigma(\bar S_0 e^{V_R})}{S_0} 
\to e^{-2\Lambda} \cR_g,
\quad \Sigma_g(\bar \cR) \equiv \Sigma(\bar \cR_g e^{-2V_R}) 
\to e^{2\Lambda} \Sigma_g(\bar \cR),
\end{eqnarray}
where the chiral projection $\Sigma$ needs to be modified 
so that the operations is covariant under the gauge symmetry. 
In the rest of this section, 
we omit the suffix $g$ attached to $\cR_g, \Sigma_g$. 
Interestingly, the U(1) gauge symmetry 
under which the compensator is charged 
becomes a gauged R-symmetry~\cite{Ferrara:1983dh}. 
We call it a U(1)$_R$ symmetry in the following discussion. 
Here, however, we do not introduce a kinetic term for $V_R$ 
and thus the vector superfield $V_R$ is an auxiliary superfield,
which should be written as a composite field 
consisting of curvature superfields $\cR$ and $\Sigma(\bar{\cR})$.

\subsection{Ghostbuster in pure $f(R)$ supergravity model } 
Let us introduce a U(1)$_R$ gauge symmetry 
under which $S_0$ has charge $c_{S_0}=1$. 
Since the chiral superfield $\cR$ = $\Sigma(\bar{S}_0)/S_0$, 
the charge of $\cR$ is determined as $c_{\cR}=-2$. 
Analogously, we find that $c_{\Sigma(\bar{\cR})}=2$. 
Then the gauged extension of the system~(\ref{fR1}) 
with $\Omega=\Omega(S,\bar S)$ is described by the action
\begin{align}
S = \left[ - \frac{3}{2} \, S_0 \, e^{V_R} \bar{S}_0 \, 
\Omega \left( \frac{\cR}{S_0} \,, \frac{\bar{\cR}}{\bar{S}_0} e^{-3V_R}\right) \right]_D 
+ \left[ S_0^3 \, \cf \left( \frac{\cR}{S_0} \,, \frac{\Sigma(\bar{\cR})}{S_0^2} \right) \right]_F,\label{originalgb}
\end{align}
where $\Omega$ should be gauge invariant
and $\cf$ should have gauge charge $c_\cf=-3$ in total. 
Hence $\cf$ should take the form
\begin{align}
\cf \left( \frac{\cR}{S_0} \,, \frac{\Sigma(\bar{\cR})}{S_0^2} \right)
\, =~ 3 \tilde{\cf} \left( \frac{\Sigma(\bar{\cR})}{S_0^2} \right) 
\frac{\cR}{S_0}.
\end{align}

To discuss the ghost elimination, 
it is useful to consider the dual system 
as in the non-gauged case~(\ref{fR2}). 
The dual system of the gauged model is 
described by 
\begin{align}
S' = & \Bigg[ -\frac{3}{2} \, S_0 \, e^{V_R} \bar{S}_0 \,
\Omega(S \,, \bar{S}e^{-3V_R}) \Bigg]_D
+ \Bigg[ 3 S_0^3 \, \tilde{\cf}(X )S \Bigg]_F \nonumber \\
& \hspace{5mm} + \Bigg[ 3 S_0^3 \, T \left( \frac{\cR}{S_0} - S \right) \Bigg]_F
+ \Bigg[ 3 S_0^3 \,Y \left( \frac{\Sigma( \bar{S}_0 \bar{S} )}{S_0^2} 
- X \right) \Bigg]_F, 
\label{eq:PureGravityGeneral}
\end{align}
where the gauge charges of $T,S,X,Y$ are 
$(c_T,c_S,c_X,c_Y)=(0 \,, -3 \,, 0 \,, -3)$. 
Similarly to the non-gauged case, we can rewrite this action as 
\begin{align}
S' \ =& ~ \Bigg[ - \frac{3}{2} \, S_0 \, e^{V_R} \bar{S}_0 
\left \{ T + \bar{T} + Y \bar{S}e^{-3V_R} +\bar{Y} e^{-3V_R} S 
+ \Omega(S, \bar{S} e^{-3V_R}) \right\} \Bigg]_D \nn \\
+& ~ \Bigg[ 3 S_0^3 \left( \tilde{\cf}(X) S - TS - XY \right) \Bigg]_F.
\end{align}
For simplicity, in the following discussion, we choose the function 
$\Omega=\gamma-h S\bar{S} e^{-3V_R}$, where
$\gamma$ is a real constant. Note that one can perform the following procedure with a more general form of $\Omega$ in a similar way. Then we obtain 
\begin{align}
S \, =& ~ \Bigg[ -\frac{3}{2} S_0 \, e^{V_R} \bar{S}_0
\left(\gamma+T+\bar{T} 
+ \left(Y\bar{S} + \bar{Y}S - h S \bar S \right) e^{-3V_R} 
\right) \Bigg]_D \nonumber \\
-& ~ \Bigg[ 3 S_0^3 \left( \tilde{\cf}(X) S + TS + XY \right) \Bigg]_F.
\end{align}
We stress that the U(1)$_R$ charges of $(S,Y)$ are 
automatically determined to be non-zero. 
This is a nontrivial and important 
nature of the $f(R)$ SUGRA model 
since the ghostbuster mechanism does not work 
if $S$ and $Y$, either of which corresponds to the ghost mode, 
did not have the U(1)$_R$ charges.

The variation of $V_R$ gives the following E.O.M for $V_R$
\begin{align}
(\gamma + T + \bar{T} ) e^{V_R} - 
2 \left( Y \bar S + S \bar Y - h S\bar S \right) e^{-2V_R} = 0.
\end{align}
This equation can be algebraically solved in terms of $V_R$ as
\begin{align}
e^{-3V_R} = \frac{\gamma + T + \bar{T}}{2 \left( Y \bar S + S \bar Y - h S \bar S \right)}.
\end{align}
Substituting this solution to the action, one finds
\begin{align}
S \, =& ~ \left[ - \frac{3}{2} S_0 \bar{S}_0 \, 
( \gamma + T + \bar{T} )^{\frac{2}{3}}
\left( Y \bar S + S \bar Y - h S \bar S \right)^{\frac{1}{3}}\right]_D \nonumber \\
-& ~ \left[ \frac{2}{\sqrt{3}} S_0^3 
\left(\tilde{\cf}(X) S + T S + X Y \right) \right]_F,\label{dualmatter}
\end{align}
where we have rescaled $S_0$ as 
$S_0 \to 2^{1/3}/\sqrt{3} \, S_0$. Thus, starting from the modified higher-curvature action~\eqref{originalgb}, we find the dual matter-coupled system~\eqref{dualmatter}.
After partial gauge fixings of superconformal symmetry\footnote{More specifically, we fix dilatiation, chiral U(1) symmetry, S-SUSY, and conformal boost, so that Poincare SUSY remains in the resultant system. The detailed procedure of superconformal gauge fixing is discussed e.g. in \cite{Freedman:2012zz}.}, 
this system becomes Poincar\'e SUGRA 
with the following K\"ahler and superpotentials,
\begin{align}
K =& -2\log \left( \gamma+T+\bar{T} \right) 
- \log \left( Y \bar S + S \bar Y - h S \bar S \right), \\
W =& -\frac{2}{\sqrt{3}} \left( \tilde{\cf}(X) S + T S + X Y \right).
\end{align}
This system is invariant 
under the U(1)$_R$ gauge transformation 
$\{ S,Y \} \to \{ e^{\Lambda}S , e^{\Lambda} Y \}$.
Therefore,  if the lowest component of $Y$ takes a non-zero value, 
we can fix the U(1)$_R$ gauge by setting $Y=1$.
Then, after a redefinition $S\to S+\frac1h$, we obtain
\begin{align}
K = & -2 \log (\gamma + T + \bar{T}) - \log(1 - h^2 S \bar S), \\
W = & - \frac{2}{\sqrt{3}} \left( \tilde{\cf}(X) (S + 1/h) + T (S+1/h) 
+ X \right).
\end{align}
If $S \not= 0$, we can also fix the gauge by setting $S=1$.
Then we find
\begin{align}
K =& -2\log (\gamma + T + \bar{T}) - \log(Y + \bar Y - h), 
\label{eq:SfixK}\\
W =& -\frac{2}{\sqrt{3}} \left(\tilde{\cf}(X) + T + X Y \right).
\end{align}
Except for the two points 
$S=0\ (Y=\infty)$, $Y=0\ (S=\infty)$, 
the above two descriptions are equivalent 
and related by a coordinate transformation between $S$ and $Y$.
In both cases, 
all the eigenvalues of the K\"ahler metric are obviously positive.
Therefore, we have shown that 
the ghost mode is eliminated by our ghostbuster mechanism. 
Note that $X$ is an auxiliary field in this setup, 
and we need to solve the E.O.M for $X$ 
to obtain the physical superpotential.

We stress that the elimination of the ghost mode 
by the ghostbuster mechanism 
in this higher-curvature system is nontrivial 
since we do not have any choice of 
the charge assignment to the superfields. 
As we have seen above, 
the would-be ghost modes have charges under U(1)$_R$, 
which enables us to remove the ghost mode 
by the gauge degree of freedom.

\subsection{Instability of scalar potential}
In this section, we analyze the scalar potential of 
the ghost-free system derived in the previous section. 
The F-term scalar potential in the Poincar\'e SUGRA is given by
\begin{eqnarray}
V = e^K \left[ K^{A\bar B} ( W_A + K_A W ) ( \bar W_{\bar B} + K_{\bar B} \bar W ) - 3|W|^2 \right].
\end{eqnarray} 
If we choose the gauge fixing condition $S=1$, 
${\rm Im} \, T$ appears only in $W$ 
due to the shift symmetry of ${\rm Im} \, T$ in the K\"ahler potential, 
and hence the mass of ${\rm Im} \, T$ is given by
\begin{eqnarray}
m_{{\rm Im} \, T}^2 \propto e^{K} ( K^{A \bar B} K_A K_{\bar B } -3 ).
\end{eqnarray}
The K\"ahler potential  in Eq.(\ref{eq:SfixK}) has the property
called the no-scale relation
\begin{eqnarray}
K^{A \bar B}K_A K_{\bar B}=3.
\end{eqnarray}     
Since $W \propto T + XY$, 
the potential has the following linear term of ${\rm Im} \, T$
\begin{eqnarray}
{\rm Im}( K_{\bar B} K^{\bar B A} W_A) \, {\rm Im} \, T \label{ImT}.
\end{eqnarray}
To realize a stable vacuum at ${\rm Im} \, T = 0$, 
this quantity must vanish identically.
By  
using the K\"ahler potential in Eq.~(\ref{eq:SfixK}), 
we find that the coefficient of the linear term is given by 
\begin{eqnarray}
{\rm Im}( K_{\bar B} K^{\bar B A} W_A ) \ = \ 
\frac{4}{\sqrt{3}} ( Y + \bar Y - h ) \, {\rm Im} X.
\end{eqnarray}
Note that the non-dynamical field $X$ 
becomes a function of $Y$ after solving its E.O.M. 
${\rm Im} \, T$ has only a mass term 
$\sim \langle Y + \bar Y - h \rangle ( X_Y Y + \overline{X}_Y \bar Y ){\rm Im} \, T$, where $X_Y\equiv \langle\partial_Y( {\rm Im}X)\rangle$. 
Unfortunately, this ``off-diagonal'' contribution 
in the mass matrix leads to a tachyonic mode.\footnote{
In general, $\langle Y + \bar Y - h \rangle$ should be nonzero 
since the K\"ahler potential has $-\log( Y + \bar Y - h )$ and 
diverges for $\langle Y+\bar Y - h \rangle=0$.} 
This instability cannot be cured 
by any higher-order terms since ${\rm Im} \, T$ appears 
only in the term~\eqref{ImT}. 
Therefore, ${\rm Im} X \not = 0$ makes ${\rm Im} \,T$ unstable 
and even if there is the local minimum in ${\rm Im}X={\rm Im} \,T=0$, 
that point cannot be a local minimum, 
but must be a saddle point. 
We conclude that although the instability caused 
by ghost mode is absent thanks to the ghostbuster mechanism, 
the pure higher-curvature action has an unstable scalar potential, which does not have any stable SUSY minimum. 
In the next section, 
we consider an extension of our model to improve this point.

\section{Stable ghostbuster model with extra matter}\label{sec:unstablemodel}

\subsection{Preliminary}
As we discussed in the previous section, 
the scalar potential of our minimal model has 
no stable SUSY minimum. 
One may improve such a situation 
by various types of modifications. 
Here we take a relatively simple way; 
We introduce an additional matter field $Z$
so that the coupling between the gravitational sector 
and the additional sector stabilizes the potential.\footnote{Even in the $R^2$ model, the deformation of scalar potential of $T$ corresponding to the scalaron superfield requires an additional degree of freedom in the dual higher-curvature SUGRA action~\cite{Cecotti:2014ipa}.}
Let us assume that $Z$ carries no U(1)$_R$ charge 
so that the superpotential $W$ contains 
$TZ$ term in the $S=1$ gauge. 
Then it is possible to introduce $Z$ in the superpotential 
in such a way that the constraint for $S$ is modified as 
\begin{eqnarray}
S = \frac{\cR}{S_0} \quad \to \quad S Z = \frac{\cR}{S_0}.
\label{ModifiedS}
\end{eqnarray} 
We can also change the definition of $X$ as
\begin{eqnarray}
 X=\frac{\Sigma(\bar S_0 \bar S)}{S_0^2} \quad \to \quad 
 X=\frac{\Sigma\left(\bar S_0 \bar S \, \bar k(\bar Z,Z) \right)}{S_0^2},
\end{eqnarray}
with an arbitrary function $k(Z,\bar Z)$. 
Note that if we chose $k(Z,\bar Z) = Z$, 
then we obtain the same unstable model 
as in Sec.\,\ref{GB}
with the redefinition $S \rightarrow S'=SZ$. 
Therefore, $k(Z,\bar{Z})$ should have 
a constant term around the minimum of $Z$, 
i.e. $k(\langle Z\rangle,\langle\bar Z\rangle) \equiv c \not =0$. 
Under this modification, the dual system is given by
\begin{align}
S' \, =& ~ \Bigg[ -\frac{3}{2} \, S_0 \, e^{V_R} \bar{S}_0 \,
\Omega( S, \bar{S} e^{-3V_R} , Z, \bar Z) \Bigg]_D 
+ \Bigg[ 3 \, S_0^3 \, T \left(\frac{\cR}{S_0} - SZ \right) \Bigg]_F \nonumber \\
+& ~\Bigg[ S_0^3 S \tilde\cf(X) \Bigg]_F 
+ \Bigg[3 S_0^3 \,Y 
\left(\frac{\Sigma(\bar{S}_0 \bar{S} \, \bar k(\bar Z))}{S_0^2}
- X \right) \Bigg]_F, 
\label{eq:PureGravityGeneral}
\end{align}
which can be rewritten as
\begin{align}
S' =& ~ \Bigg[ - \frac{3}{2} \, S_0 \, e^{V_R} \bar{S}_0 
\left\{ T + \bar{T} + Y\bar{S} e^{-3V_R} \, \bar k(\bar Z) + 
\bar{Y} e^{-3V_R} S \, k(Z) +\Omega \right\} \Bigg]_D
\nonumber \\
+& ~ \Bigg[ S_0^3 \left(\tilde \cf (X)S-3TSZ-3XY\right) \Bigg]_F.
\end{align}
For simplicity, let us choose the function as 
\begin{eqnarray}
\Omega = \gamma-g(Z,\bar Z)- h( Z,\bar Z) \, S\bar{S} \, e^{-3V_R}.
\end{eqnarray}
After solving the E.O.M for $V_R$, 
we find the following K\"ahler potential and superpotential  
\begin{align}
K =& -2 \log \Big[ \gamma+T+\bar{T}-g(Z,\bar Z) \Big]
- \log \Big[ Y \bar k(\bar Z) +\bar Y k(Z) - h(Z,\bar Z) \Big], \\
W =& \, \frac{2}{\sqrt{3}} \left[ \frac{1}{3} \tilde \cf(X) - \left(T Z +X Y \right) \right],
\end{align}
in the $S=1$ gauge.

\subsection{Example of matter coupled $f(R)$ supergravity}\label{exfr}
Let us discuss a simple example by setting the functions as
\begin{eqnarray}
k(Z)=c +Z, \quad \quad \quad 
\Omega =\gamma + (\beta - b Z\bar Z) \, S \bar S \, e^{-3V_R}. \label{SimpleModel}
\end{eqnarray}
The corresponding K\"ahler potential is given by
\begin{align}
 K&=-2\log \omega_1 \label{SMK}
-\log \omega_2, \\
\omega_1&\equiv \gamma+T+\bar{T},\\
\omega_2&\equiv \beta + \left( \bar Y(c+Z)+{\rm c.c.} \right) 
- b Z \bar Z,
\end{align}
where both $\omega_1$ and $\omega_2$ 
are required to be positive so that 
there exists a solution of the E.O.M. for $V_R$ 
and the condition $e^K > 0$.
The eigenvalues $\{ \lambda_i \, | \, i=1,2,3 \}$ 
of the K\"ahler metric $K_{A\bar B}$ are given by
\begin{eqnarray}
\lambda_1 = \frac{2}{\omega_1^2}, 
\quad \lambda_2 + \lambda_3 = 
\frac{|\partial_Y \omega_2|^2 + |\partial_Z\omega_2|^2 + 
b \, \omega_2}{\omega_2^2},
\quad \lambda_2\lambda_3 = 
\frac{b |c|^2-\beta}{\omega_2^3}.
\end{eqnarray} 
Furthermore, by choosing the function $\tilde {\cal F}$ 
so that $\tilde {\cal F}(0)=0 ,\, \tilde {\cal F}'(0) =0$, 
we find a SUSY vacuum satisfying $W_A=W=0$
at $X=Y=T=S=0$, which is guaranteed to be stable.
Therefore, there exists the SUSY vacuum 
with a positive definite metric if and only if
\begin{eqnarray}
\gamma = \omega_1 |_{\rm vac} > 0, \quad 
\beta = \omega_2 |_{\rm vac} > 0, \quad 
b > \frac{\beta}{|c|^2}, \quad 
c \not= 0.
\end{eqnarray} 
When these conditions are satisfied, 
there exist no ghost anywhere in the region 
${\cal M} = \{T,Y,Z \,|\, \omega_1>0 \,,\, \omega_2>0 \}$ 
and the boundary $\partial {\cal M}$ 
is geodesically infinitely far away from the SUSY vacuum.

\section{Ghostbuster mechanism from higher-curvature SUGRA viewpoint }
\label{GBC}
In this section, we discuss how the ghostbuster mechanism works in the higher-curvature frame. As we have seen in previous two sections, the ghost supermultiplet is eliminated in both pure and matter-coupled higher-curvature systems. 

Let us consider the original action 
for $f(R)$ gravity before taking the dual transformation. For concreteness of the discussion, we take the simplest model with an additional matter superfield in Eq.~(\ref{SimpleModel}). The same conclusion follows even in the absence of an additional matter.
The higher-curvature action can be obtained 
by solving E.O.M. for $T$ and $Y$
and imposing the constraints for $S$ and $X$. 
Here we introduce $S_1 \equiv c S_0 S+\cR_g $ as an extra matter 
and solve the modified constraint (\ref{ModifiedS}) for $Z$. 
After introducing the quadratic term of $X$, 
the original action takes the form
\begin{align}
S' =& ~ \bigg[-\frac{3}{2}\gamma |S_0|^2e^{V_R}
-\frac{3\beta}{2|c|^2}  |S_1-\cR_g |^2 e^{-2V_R}+\frac32 a |S_0 X|^2e^{V_R} +\frac32 b |\cR_g|^2 e^{-2V_R}  \bigg]_D\nn \\
+& ~ \bigg[ S_0^2 (S_1-\cR_g) X {\cal G}(X) \bigg]_F
\end{align}
with $\tilde \cf(X) \equiv c X {\cal G}(X)$ and 
\begin{eqnarray}
\cR_g = \frac{\Sigma(\bar S_0 e^{V_R})}{S_0}, \quad \quad
X= \frac{\Sigma(\bar S_1 e^{-2V_R} )}{S_0^2},
\end{eqnarray}
where $a$ and $b$ are real (positive) parameters. 
Note that $X$ now does not have 
the Ricci scalar in the lowest component 
but a higher-derivative superfield made out of $S_1$. 
This means that the higher-derivative term of $\cR_g$ 
is now replaced by that of $S_1$, 
and hence the higher-curvature term does not show up. 
By expanding the action explicitly, 
one can check that this action has Ricci scalar terms 
up to the quadratic order. 
We note that, however, this does not lead to the conclusion 
that the ghost is removed by the additional matter: 
since there still exist higher-derivative terms of $S_1$, 
the ghost mode can arise from such terms. 
One may also confirm that 
the absence of the higher curvature terms $R^n \ (n\geq3)$ 
is not an artifact of field redefinition. 
We can show that in this specific matter coupled model, 
the higher-curvature terms exist only in the off-shell action 
before substituting the solution of the E.O.M 
for the auxiliary field in $V_R$.

We stress that this conclusion does not mean that the higher-curvature modification is removed by the ghostbuster mechanism. As we claimed above, the resultant system has scalar curvature terms only up to the quadratic order, as the simplest Ceccoti model does~\cite{Cecotti:1987sa}. However, the coupling of the resultant system is completely different from the Ceccoti model. In our dual matter coupled system in Sec.~\ref{exfr}, K\"ahler potential~\label{SMK} takes the form
\begin{equation}
K\sim-2\log(T+\bar{T})-\log(Y+\bar Y+\cdots),
\end{equation}
whereas, in the Ceccoti model, it can be written as
\begin{equation}
K=-3\log(T_{c}+\bar{T}_c+\cdots),
\end{equation}
where $T,Y$ and $T_c$ are chiral superfields. The difference of the K\"ahler potentials leads to a different moduli space geometry. Interestingly, all $T,Y$ and $T_c$ have the hyperbolic geometry structure, which is applicable to the so-called inflationary $\alpha$-attractors~\cite{Kallosh:2013yoa,Carrasco:2015uma}. In the $\alpha$-attractor inflation, we take the moduli space $K=-3\alpha\log (\Phi+\bar{\Phi})$ for an inflaton superfield $\Phi$, and the value of the parameter $\alpha$ has a relation to the tensor to scalar ratio $r$ as $r=\frac{12\alpha}{N^2}$, where $N$ is the number of e-foldings at the horizon exit. In our model, we have $\alpha=\frac13$ and $\frac23$, whereas the Ceccoti model has $\alpha=1$. If we apply our model to inflation, we would find a value of tensor to scalar ratio $r$ different from that of the Ceccoti model. Therefore, the higher-curvature modification has physical consequences even though the higher-order scalar curvature terms seem to disappear after the ghostbuster mechanism. Since the construction of the inflation model is beyond the scope of this paper, we leave it as future work.


\section{Conclusion}
\label{conclusion}
We have applied the ghost buster method 
to a higher-curvature system of SUGRA. 
It has been known that once we introduce 
a higher scalar curvature multiplet $\Sigma(\bar{\mathcal R} )$, 
a ghost mode generically shows up in the system 
as we reviewed in Sec.\,\ref{review}. 
The ghostbuster method requires 
a nontrivial U(1) gauge symmetry 
with a non-propagating gauge superfield. 
It turned out that the required U(1) symmetry should be 
the gauged R-symmetry in the case of the higher-curvature system, since the ghost arises from the gravitational superfield. 
Due to the uniqueness of the gauge charge assignment, 
it is nontrivial that if the ghostbuster method is applicable 
to remove the ghost. 
As we have shown in Sec.\,\ref{GB}, 
thanks to the nonzero U(1) charge of ``would-be'' ghost mode, 
we can eliminate the ghost mode and 
obtain a ghost-free action. 
However, the resultant ghost-free system 
turned out to be unstable 
because of the scalar potential instability. 
Such an instability is easily cured by introducing matter fields, 
which would be necessary for realistic models. 
Additional matter superfields can stabilize the scalar potential 
if we choose proper couplings 
between gravity and matter multiplets. 

We have also discussed how the ghostbuster mechanism can be seen in the higher-curvature system in Sec.~\ref{GBC}. We have found that the higher-order scalar curvature terms $R^n$ with $n\geq3$ are eliminated in using the mechanism, and the resultant system has the scalar curvature up to the quadratic order. However, the higher-curvature modification is not completely eliminated by the mechanism. We find moduli space geometry different from the known $R+R^2$ supergravity~\cite{Cecotti:1987sa}. Therefore, despite the absence of $f(R)$ type interactions in the final form, the SUSY higher-order curvature corrections give physical differences. In particular, the difference of the moduli space structure might be useful for constructing inflationary models.

In this work, we did not discuss 
the elimination of ghosts originated from 
higher-derivative terms of matter superfields. 
It is a straightforward extension of 
our previous work~\cite{Fujimori:2016udq} 
for global SUSY to SUGRA 
and is much easier than the higher-curvature model discussed in this paper, 
since the U(1) charge assignment is 
not unique for matter higher-derivative models. 
Since the higher-derivatives of matter fields in SUGRA 
requires the compensator $S_0$, 
it would be interesting to assign the U(1) charge 
to the compensator as well, 
i.e. we can use U(1) R-symmetry for the ghostbuster mechanism 
as with the higher-curvature case, 
which is only possible for the SUGRA case. 

Let us mention the applicability of our mechanism 
to the other SUGRA formulations, 
where the auxiliary fields in the gravity multiplet are different. 
Our mechanism is not applicable 
for the so-called new minimal SUGRA formulation~\cite{Sohnius:1981tp}, 
since the compensator is a real linear superfield, 
which cannot have any U(1) charge. 
For the non-minimal SUGRA case, 
it would be possible to assign a nontrivial U(1) charge 
to complex linear compensator. 
In addition, it is known that 
the $R^2$ model of non-minimal SUGRA 
has a ghost mode in the spectrum, 
so it is interesting to see 
if the ghost can be removed by our mechanism.
\section*{Acknowledgement}
This work is supported by  the Ministry of Education,
Culture, Sports, Science (MEXT)-Supported Program for the Strategic
Research Foundation at Private Universities ``Topological Science''
(Grant No.~S1511006).
The work of M.~N.~is also supported in part by a Grant-in-Aid for
Scientific Research on Innovative Areas ``Topological Materials
Science'' (KAKENHI Grant No.~15H05855) from the MEXT of Japan, and 
by the Japan Society for the Promotion of Science
(JSPS) Grant-in-Aid for Scientific Research (KAKENHI Grant
No.~16H03984).
Y.~Y.~is supported by SITP and by the NSF Grant PHY-1720397.
\appendix

\section{Superconformal tensor calculus}
\label{Components}
Here we give a brief summary of the superconformal formulation. 
We use the convention 
$\eta_{ab}={\rm diag}(-1,1,1,1)$ for the Minkowski metric.

In 4D ${\cal N}=1$ conformal SUGRA, 
we have the super-Poincar\'e generators 
$\{ P_a, M_{ab} Q_\alpha \}$, 
and the additional superconformal generators, 
$\{{\bf D},{\bf A},S_\alpha, K_a\}$. 
They correspond to the translation $P_a$, 
the Lorentz rotation $M_{ab}$, 
the SUSY $Q_\alpha$, 
the dilatation ${\bf D}$, 
the chiral U(1) ${\bf A}$, 
the S-SUSY $S_\alpha$ and 
the conformal boost $K_a$, respectively. 
Such additional gauge degrees of freedom are 
technically useful for the construction of the SUGRA action. 
In conformal SUGRA, a supermultiplet is characterized 
by the charges under ${\bf D}$ and ${\bf A}$ 
denoted by $w$ and $n$, respectively. 
We introduce one particular supermultiplet called the compensator, 
whose components are auxiliary fields 
or removed by the superconformal gauge fixing. 
In this paper, we use a chiral superfield as the compensator, 
which gives the so-called old-minimal SUGRA 
after the superconformal gauge fixing.

In the following, 
we summarize the component expressions of supermultiplets, 
the chiral projection operation, the invariant formulae 
and some identities.
\paragraph{General multiplet \\}
The components of a general multiplet with charges $(w,n)$ 
are given by 
\begin{eqnarray}
 {\cal C} \, = \, (C,\zeta,H,K,B_a,\lambda,D) \, \in \, {\cal G}_{(w,n)}, \label{GM}
\end{eqnarray}
whose charge conjugate $\bar {\cal C}$ is
\begin{eqnarray}
 \bar {\cal C} \, = \, (C^*,\zeta^c,H^*,K^*,B_a^*,\lambda^c,D^*) \, \in \, {\cal G}_{(w,-n)},
\end{eqnarray}
where $\zeta^c$ and $\lambda^c$ are 
the charge conjugates of $\zeta$ and $\lambda$, respectively. 
Note that $\bar {\cal C}$ has the ${\bf D}$ and ${\bf A}$ charges $(w,-n)$.
\paragraph{Multiplication law \\}
Here We show the multiplication rule of supermultiplets. 
Suppose ${\cal C}^I \in {\cal G}_{(w_I,n_I)}$ 
and consider a function $f({\cal C}^I) \in {\cal G}_{(w,n)}$. 
The component of $f({\cal C}^I)$ is given by
\begin{eqnarray}
f({\cal C}^I) = \bigg[
f(C^I) \,,\, f_I \zeta^I ,\, f_I H^I + \cdots ,\, f_I K^I + \cdots ,\, 
f_I B^I_a+\cdots,\, f_I \lambda^I + \cdots , \nonumber \\
f_I D^I + \frac{1}{2} f_{IJ}
\left( H^I H^J + K^I K^J - B_a^I B^{Ja} - D_a C^I D^a C^J \right) + \cdots \bigg],
\end{eqnarray}
where ellipses denote terms containing fermions 
and $f_I, f_{IJ}$ are derivatives defined as 
\begin{eqnarray}
f_I\equiv \frac{\partial f(C)}{\partial C^I},\quad 
f_{IJ}\equiv \frac{\partial^2 f(C)}{\partial C^I \partial C^J},\quad 
\end{eqnarray}
and the covariant derivative of $C^I$ is given by
\begin{eqnarray}
D_a C = e_a{}^\mu ( \partial_\mu - \omega b_\mu - in A_\mu ) C +\cdots.
\end{eqnarray}
Note that since $(w,n)$ are additive quantum numbers, 
the following relations are satisfied,
\begin{eqnarray}
\sum_{I} w_I f_I C^I = w f(C), \quad 
\sum_{I} n_I f_I C^I = n f(C), \quad 
\sum_{J} w_J f_{IJ} C^J = (w-w_I) f_I(C).
\end{eqnarray}

\paragraph{Chiral multiplet and chiral projection \\}
The components of a chiral superfield are given by
\begin{eqnarray}
\Phi = ( \phi , P_L \chi , F ) \in \Sigma_{w},
\end{eqnarray}
where $P_L=\frac{{\bf 1}+\gamma_5}2$ is 
the chirality projection operator. 
A chiral (anti-chiral) multiplet satisfies the constraint $w=n (-n)$
and can be embedded 
into a general multiplet \eqref{GM} as
\begin{eqnarray}
\Phi ~ \to ~ {\cal C}(\Phi) = 
\left( \phi \,, -\sqrt{2}i P_L \chi \,, -F \,,\, iF \,, i D_a \phi \,, 0 \,, 0 \right) 
\in {\cal G}_{(w,w)},
\end{eqnarray}
whereas an anti-chiral multiplet 
$\bar \Phi = (\phi^*, P_R \chi \,, F^* ) \in \bar\Sigma_{w}$ 
can be embedded as 
\begin{eqnarray}
\bar \Phi ~ \to ~ {\cal C}(\bar\Phi) = 
\left( \bar\phi \,,\sqrt{2}i P_R \chi \,, -\bar F \,,-i\bar F \,, 
-i D_a \bar\phi \,, 0 \,, 0 \right) \in {\cal G}_{(w,-w)}.
\end{eqnarray}
 
One can make a chiral multiplet 
out of a general multiplet satisfying $w-n=2$, 
and we refer to this operation 
as the chiral projection $\Sigma$ 
\begin{eqnarray}
\Sigma: ~ {\cal C}\in {\cal G}_{(w,w-2)} ~ \to ~
\Sigma({\cal C})\in \Sigma_{w+1}, 
\end{eqnarray}
whose components are given by
\begin{eqnarray}
\Sigma({\cal C}) = 
\left[
\frac12(H-iK) \,,\, 
\frac{i}{\sqrt{2}} P_L ( \lambda + \gamma^a D_a \zeta ) \,,\,
-\frac{1}{2} (D + D^a D_a C + iD^a B_a ) 
\right]. 
\end{eqnarray}
where
\begin{eqnarray}
D^a D_a C &=& 
e^{\mu a} \left( \partial_\mu - (w+1) b_\mu - i n A_\mu \right )
D_a C -\omega_a{}^{ab} D_b C+2w f_a{}^a C+\cdots, \\
D^a B_a &=& e^{a\mu} ( \partial_\mu - (w+1) b_\mu 
- i n A_\mu ) B_a - \omega_a{}^{ab} B_b + 2 in f_a^a C + \cdots.
\end{eqnarray}
Here the ellipses denote terms containing fermions, which we do not focus on in this paper.
In particular, for ${\cal C}\in {\cal G}_{(2,0)}$, we find that
\begin{eqnarray}
\hspace{-5mm} 
D^a D_a C + i D^a B_a + {\rm c.c.} = - \frac{1}{3} R ( C + \bar C ) +
e^{-1} \partial_\mu \left( e e^{\mu a} ( D_a C + i B_a + {\rm c.c.} ) \right) + \cdots, 
\end{eqnarray}
where we have used 
\begin{eqnarray}
\omega_b{}^{ba} = -3 b^a - e^{-1} \partial_\mu ( e e^{\mu a} ) + \cdots ,\quad f_a{}^a = -\frac{1}{12}R + \cdots .
\end{eqnarray}
For instance, we can construct a chiral superfield 
out of a chiral and an anti-chiral superfield:
\begin{eqnarray}
\Phi \in \Sigma_0, \quad S_0 \in \Sigma_1, \quad \to \quad 
\Phi' = \frac{\Sigma(\bar S_0 \bar \Phi)}{S_0^2}  \in \Sigma_0. 
\end{eqnarray} 
Note that the chiral projection does not act on a chiral multiplet, 
i.e. for ${\cal C} \in {\cal G}_{n,n-2}, \ \Phi \in \Sigma_{m}$, 
we find that
\begin{eqnarray}
\Sigma({\cal C}\Phi) =  \Sigma({\cal C})\Phi   \in \Sigma_{n+m+1}.
\end{eqnarray}

\paragraph{Vector multiplet ${\cal V}\in {\cal G}_{(0,0)}$ 
and gauge transformation \\}
We define a gauge vector superfield as
\begin{eqnarray}
V \in {\cal V}: \quad V \in {\cal G}_{(0,0)}, \quad V=\bar V
\end{eqnarray}
The composite supermultiplet $\bar \Phi \, e^{2g V} \Phi$
is invariant under the SUSY gauge transformation 
with $\Lambda \in \Sigma_0$,
\begin{eqnarray}
e^{2gV} \to ~ e^{2gV'} = e^{-g \bar \Lambda} \, e^{2gV} e^{-g \Lambda},\quad 
\Phi \to \Phi' =e^{g \Lambda}\Phi,\quad 
\bar \Phi \to \bar \Phi' =\bar \Phi e^{g \bar \Lambda}.
\end{eqnarray}
Under this transformation, 
a chiral supermultiplet 
$\tilde \Phi \equiv \Sigma( \bar \Phi e^{2g V} )$ transforms as  
\begin{eqnarray}
\tilde \Phi \quad \to\quad 
\tilde \Phi' 
= \Sigma\left(\bar \Phi e^{2g V} e^{-g\Lambda}\right)
= \tilde \Phi e^{-g \Lambda }.
\end{eqnarray}
We take the Wess-Zumino gauge, 
in which the components of $V$ are given by
\begin{eqnarray}
V|_{\rm WZ}=[B_\mu^g, \lambda^g, D^g], \quad 
{\cal C}(V|_{\rm WZ})=[0,0,0,0,B_a^g,\lambda^g,D^g].
\end{eqnarray}
Here the ordinary gauge transformation 
with $\Lambda=[i\theta,0,0]$ is given by
\begin{eqnarray}
 B_\mu^{'g} =B^g_\mu+\partial_\mu \theta, \quad  \Phi'=e^{ig\theta }\Phi.
\end{eqnarray}
\paragraph{Invariant action formula \\}
The superconformal invariant actions are 
given by the $D$- and $F$-term density formulas:
the $D$-term invariant formula 
with ${\cal C} \in {\cal G}_{(2,0)}$ and 
$\bar{\cal C} ={\cal C}$ is given by
\begin{eqnarray}
 [{\cal C}]_{D} \equiv \int d^4x \, e 
 \left(D - \frac{1}{3} R  C +\cdots\right),
\end{eqnarray}
and the $F$-term invariant formula with $\Phi \in \Sigma_3$ is
\begin{eqnarray}
[\Phi]_F \equiv \int d^4x \,e ( F + \cdots ) + {\rm c.c.}. 
\end{eqnarray}
There is a useful identity 
between these two invariant formulas
\begin{eqnarray}
\left[ \Sigma({\cal C}) \right]_F 
= -\frac{1}{2} [ {\cal C} + \bar{\cal C}]_D
\quad {\rm for~} {\cal C}\in {\cal G}_{(2,0)}.
\end{eqnarray}
Then, for $T, X \in \Sigma_1$,
\begin{eqnarray}
\left[ \Sigma(\bar X) T \right]_F
= \left[ \Sigma(\bar X T) \right]_F
= - \frac{1}{2} \left[ \bar X T + \bar T X \right]_D. \label{ID}
\end{eqnarray}

In addition, for U(1) charged chiral multiplets 
$\Phi, \tilde \Phi \in \Sigma_1$,
we find that
\begin{eqnarray}
\left[ \Sigma \left(\bar \Phi e^{2g V} \right) \! \tilde \Phi \right]_F
=-\frac12 \left[ \bar \Phi e^{2g V} \tilde \Phi + {\rm c.c.} \right]_D.
\end{eqnarray}

\paragraph{Composite supermultiplets \\}
We finally show the components of composite superfields.
With $\bar{\Phi} \in \bar{\Sigma}_1$, 
we can make a chiral superfield with $w=n=2$ as
\begin{eqnarray}
\Sigma(\bar \Phi) = \left[ -\bar F \,,\, P_L \gamma^aD_a \chi \,,\, 
-D_a D^a \bar \phi \right] \in \Sigma_{2}.
\end{eqnarray}
The composite anti-chiral superfield 
$\bar \Phi \, e^{2c V} \in \bar \Sigma_2$ 
can be embedded into a general multiplet as
\begin{eqnarray}
{\cal C}( \bar \Phi \, e^{2c V} ) \big|_{\rm WZ} =
\Big[ \bar \phi \,,\, \sqrt{2} i P_R \chi \,,\, -\bar F \,,\, -i\bar F \,,\, 
- i D_a \bar \phi + 2 c B_a^g 
\bar \phi +\cdots , \nonumber \\
\cdots ,\, 2c D^g \bar \phi + 2 i c B^g_a D^a \bar \phi 
- 2 \bar \phi c^2 B_a^g B^{ga} \Big].
\end{eqnarray}
Note that $c$ denotes a gauge charge of $\Phi$ under $V$.
In addition, the projected composite superfield takes the form
\begin{eqnarray}
\Sigma(\bar \Phi \, e^{2cV}) \big|_{\rm WZ} = 
\left[ -\bar F \,,\, \cdots,\, -c D^g \bar \phi - {\cal D}^a {\cal D}_a \bar \phi \right],
\end{eqnarray}
with 
\begin{eqnarray}
{\cal D}_\mu \bar \phi = D_\mu \bar \phi + i c B_\mu^g \bar \phi.
\end{eqnarray}



\begin{thebibliography}{99}

\bibitem{Fujimori:2016udq} 
  T.~Fujimori, M.~Nitta and Y.~Yamada,
  ``Ghostbusters in higher derivative supersymmetric theories: who is afraid of propagating auxiliary fields?,''
  JHEP {\bf 1609}, 106 (2016)
  [arXiv:1608.01843 [hep-th]].
  
\bibitem{Ostrogradski}
 M. Ostrogradski, 
 ``Memoires sur les equations differentielles relatives au probleme des
  isoperimetres,'' 
 Mem. Ac. St. Petersbourg VI 4, 385 (1850).
 
\bibitem{Woodard:2006nt} 
  R.~P.~Woodard,
  ``Avoiding dark energy with 1/r modifications of gravity,''
  Lect.\ Notes Phys.\  {\bf 720}, 403 (2007)
  [astro-ph/0601672].


\bibitem{Horndeski:1974wa} 
  G.~W.~Horndeski,
  ``Second-order scalar-tensor field equations in a four-dimensional space,''
  Int.\ J.\ Theor.\ Phys.\  {\bf 10}, 363 (1974).


\bibitem{Kobayashi:2011nu} 
  T.~Kobayashi, M.~Yamaguchi and J.~Yokoyama,
  ``Generalized G-inflation: Inflation with the most general second-order field equations,''
  Prog.\ Theor.\ Phys.\  {\bf 126}, 511 (2011)
  [arXiv:1105.5723 [hep-th]].

\bibitem{Nicolis:2008in} 
  A.~Nicolis, R.~Rattazzi and E.~Trincherini,
  ``The Galileon as a local modification of gravity,''
  Phys.\ Rev.\ D {\bf 79}, 064036 (2009)
  [arXiv:0811.2197 [hep-th]].
  

\bibitem{Farakos:2013zya} 
  F.~Farakos, C.~Germani and A.~Kehagias,
  ``On ghost-free supersymmetric galileons,''
  JHEP {\bf 1311}, 045 (2013)
  [arXiv:1306.2961 [hep-th]].


\bibitem{Roest:2017uga} 
  D.~Roest, P.~Werkman and Y.~Yamada,
  ``Internal Supersymmetry and Small-field Goldstini,''
  arXiv:1710.02480 [hep-th].


\bibitem{Starobinsky:1980te} 
  A.~A.~Starobinsky,
  ``A New Type of Isotropic Cosmological Models Without Singularity,''
  Phys.\ Lett.\  {\bf 91B}, 99 (1980).


\bibitem{Ade:2015lrj} 
  P.~A.~R.~Ade {\it et al.} [Planck Collaboration],
  ``Planck 2015 results. XX. Constraints on inflation,''
  Astron.\ Astrophys.\  {\bf 594}, A20 (2016)
  [arXiv:1502.02114 [astro-ph.CO]].


\bibitem{Buchdahl:1983zz} 
  H.~A.~Buchdahl,
  ``Non-linear Lagrangians and cosmological theory,''
  Mon.\ Not.\ Roy.\ Astron.\ Soc.\  {\bf 150}, 1 (1970).


\bibitem{DeFelice:2010aj} 
  A.~De Felice and S.~Tsujikawa,
  ``f(R) theories,''
  Living Rev.\ Rel.\  {\bf 13}, 3 (2010)
  [arXiv:1002.4928 [gr-qc]].
\bibitem{Nojiri:2017ncd} 
  S.~Nojiri, S.~D.~Odintsov and V.~K.~Oikonomou,
  ``Modified Gravity Theories on a Nutshell: Inflation, Bounce and Late-time Evolution,''
  Phys.\ Rept.\  {\bf 692}, 1 (2017)
  [arXiv:1705.11098 [gr-qc]].
  
\bibitem{Gates:1995fx-0} 
  S.~J.~Gates, Jr.,
  ``Why auxiliary fields matter: The Strange case of the 4-D, N=1 supersymmetric QCD effective action,''
  Phys.\ Lett.\ B {\bf 365}, 132 (1996)
  [hep-th/9508153].
\bibitem{Gates:1995fx-1}
  S.~J.~Gates, Jr.,
  ``Why auxiliary fields matter: The strange case of the 4-D, N=1 supersymmetric QCD effective action. 2.,''
  Nucl.\ Phys.\ B {\bf 485}, 145 (1997)
  [hep-th/9606109].
  
  


\bibitem{Antoniadis:2007xc} 
  I.~Antoniadis, E.~Dudas and D.~M.~Ghilencea,
  ``Supersymmetric Models with Higher Dimensional Operators,''
  JHEP {\bf 0803}, 045 (2008)
  [arXiv:0708.0383 [hep-th]].



  \bibitem{Dudas:2015vka} 
  E.~Dudas and D.~M.~Ghilencea,
    ``Effective operators in SUSY, superfield constraints and searches for a UV completion,''
  JHEP {\bf 1506}, 124 (2015)
  [arXiv:1503.08319 [hep-th]].
  
\bibitem{Khoury:2010gb} 
  J.~Khoury, J.~L.~Lehners and B.~Ovrut,
  ``Supersymmetric P(X,$\phi$) and the Ghost Condensate,''
  Phys.\ Rev.\ D {\bf 83}, 125031 (2011)
  [arXiv:1012.3748 [hep-th]].

\bibitem{Khoury:2011da} 
  J.~Khoury, J.~L.~Lehners and B.~A.~Ovrut,
  ``Supersymmetric Galileons,''
  Phys.\ Rev.\ D {\bf 84}, 043521 (2011)
  [arXiv:1103.0003 [hep-th]].

\bibitem{Koehn:2012ar} 
  M.~Koehn, J.~L.~Lehners and B.~A.~Ovrut,
  ``Higher-Derivative Chiral Superfield Actions Coupled to N=1 Supergravity,''
  Phys.\ Rev.\ D {\bf 86}, 085019 (2012)
  [arXiv:1207.3798 [hep-th]].

\bibitem{Koehn:2012te} 
  M.~Koehn, J.~L.~Lehners and B.~Ovrut,
  ``Ghost condensate in $N=1$ supergravity,''
  Phys.\ Rev.\ D {\bf 87}, no. 6, 065022 (2013)
  [arXiv:1212.2185 [hep-th]].


\bibitem{Fujimori:2017kyi} 
  T.~Fujimori, M.~Nitta, K.~Ohashi, Y.~Yamada and R.~Yokokura,
  ``Ghost-free vector superfield actions in supersymmetric higher-derivative theories,''
  JHEP {\bf 1709}, 143 (2017)
  [arXiv:1708.05129 [hep-th]].


\bibitem{Buchbinder:1994iw-0} 
  I.~L.~Buchbinder, S.~Kuzenko and Z.~Yarevskaya,
  ``Supersymmetric effective potential: Superfield approach,''
  Nucl.\ Phys.\ B {\bf 411}, 665 (1994).
\bibitem{Buchbinder:1994iw-1}
  I.~L.~Buchbinder, S.~M.~Kuzenko and A.~Y.~Petrov,
  ``Superfield chiral effective potential,''
  Phys.\ Lett.\ B {\bf 321}, 372 (1994).
\bibitem{Buchbinder:1994iw-2}
  A.~T.~Banin, I.~L.~Buchbinder and N.~G.~Pletnev,
  ``On quantum properties of the four-dimensional generic chiral superfield model,''
  Phys.\ Rev.\ D {\bf 74}, 045010 (2006)
  [hep-th/0606242].
\bibitem{Buchbinder:1994iw-3}
  S.~M.~Kuzenko and S.~J.~Tyler,
  ``The one-loop effective potential of the Wess-Zumino model revisited,''
  JHEP {\bf 1409}, 135 (2014)
  [arXiv:1407.5270 [hep-th]].


\bibitem{Nitta:2014fca} 
  M.~Nitta and S.~Sasaki,
  ``Higher Derivative Corrections to Manifestly Supersymmetric Nonlinear Realizations,''
  Phys.\ Rev.\ D {\bf 90}, no. 10, 105002 (2014)
  [arXiv:1408.4210 [hep-th]].


\bibitem{Farakos:2012qu} 
  F.~Farakos and A.~Kehagias,
  ``Emerging Potentials in Higher-Derivative Gauged Chiral Models Coupled to N=1 Supergravity,''
  JHEP {\bf 1211}, 077 (2012)
  [arXiv:1207.4767 [hep-th]].


\bibitem{Sasaki:2012ka} 
  S.~Sasaki, M.~Yamaguchi and D.~Yokoyama,
  ``Supersymmetric DBI inflation,''
  Phys.\ Lett.\ B {\bf 718}, 1 (2012)
  [arXiv:1205.1353 [hep-th]].

\bibitem{Aoki:2014pna-0} 
  S.~Aoki and Y.~Yamada,
  ``Inflation in supergravity without K\"ahler potential,''
  Phys.\ Rev.\ D {\bf 90}, no. 12, 127701 (2014)
  [arXiv:1409.4183 [hep-th]].
\bibitem{Aoki:2014pna-1} 
  S.~Aoki and Y.~Yamada,
  ``Impacts of supersymmetric higher derivative terms on inflation models in supergravity,''
  JCAP {\bf 1507}, no. 07, 020 (2015)
  [arXiv:1504.07023 [hep-th]].
  
\bibitem{Adam:2013awa-0} 
  C.~Adam, J.~M.~Queiruga, J.~Sanchez-Guillen and A.~Wereszczynski,
  ``Extended Supersymmetry and BPS solutions in baby Skyrme models,''
  JHEP {\bf 1305}, 108 (2013)
  [arXiv:1304.0774 [hep-th]].

\bibitem{Adam:2013awa-1} 
  C.~Adam, J.~M.~Queiruga, J.~Sanchez-Guillen and A.~Wereszczynski,
  ``N=1 supersymmetric extension of the baby Skyrme model,''
  Phys.\ Rev.\ D {\bf 84}, 025008 (2011)
  [arXiv:1105.1168 [hep-th]].

\bibitem{Nitta:2014pwa-0} 
  M.~Nitta and S.~Sasaki,
  ``BPS States in Supersymmetric Chiral Models with Higher Derivative Terms,''
  Phys.\ Rev.\ D {\bf 90}, no. 10, 105001 (2014)
  [arXiv:1406.7647 [hep-th]].
\bibitem{Nitta:2014pwa-1} 
  M.~Nitta and S.~Sasaki,
  ``Classifying BPS States in Supersymmetric Gauge Theories Coupled to Higher Derivative Chiral Models,''
  Phys.\ Rev.\ D {\bf 91}, 125025 (2015)
  [arXiv:1504.08123 [hep-th]].

\bibitem{Bolognesi:2014ova-0} 
  S.~Bolognesi and W.~Zakrzewski,
  ``Baby Skyrme Model, Near-BPS Approximations and Supersymmetric Extensions,''
  Phys.\ Rev.\ D {\bf 91}, no. 4, 045034 (2015)
  [arXiv:1407.3140 [hep-th]].
\bibitem{Bolognesi:2014ova-1} 
  J.~M.~Queiruga,
  ``Baby Skyrme model and fermionic zero modes,''
  Phys.\ Rev.\ D {\bf 94}, no. 6, 065022 (2016)
  [arXiv:1606.02869 [hep-th]].

\bibitem{Gudnason:2015ryh-0} 
  S.~B.~Gudnason, M.~Nitta and S.~Sasaki,
  ``A supersymmetric Skyrme model,''
  JHEP {\bf 1602}, 074 (2016)

  [arXiv:1512.07557 [hep-th]].
\bibitem{Gudnason:2015ryh-1} 
  S.~B.~Gudnason, M.~Nitta and S.~Sasaki, 
  ``Topological solitons in the supersymmetric Skyrme model,''
  JHEP {\bf 1701}, 014 (2017)
  [arXiv:1608.03526 [hep-th]].

\bibitem{Queiruga:2015xka} 
  J.~M.~Queiruga,
  ``Skyrme-like models and supersymmetry in 3+1 dimensions,''
  Phys.\ Rev.\ D {\bf 92}, no. 10, 105012 (2015)
  [arXiv:1508.06692 [hep-th]].

\bibitem{Queiruga:2017blc}
  J.~M.~Queiruga and A.~Wereszczynski,
  ``Non-uniqueness of the supersymmetric extension of the $O(3)$ $\sigma$-model,''
  JHEP {\bf 1711}, 141 (2017)
  [arXiv:1703.07343 [hep-th]].

\bibitem{Eto:2012qda} 
  M.~Eto, T.~Fujimori, M.~Nitta, K.~Ohashi and N.~Sakai,
  ``Higher Derivative Corrections to Non-Abelian Vortex Effective Theory,''
  Prog.\ Theor.\ Phys.\  {\bf 128}, 67 (2012)
  [arXiv:1204.0773 [hep-th]].

\bibitem{Nitta:2017mgk} 
 M.~Nitta, S.~Sasaki and R.~Yokokura,
 ``Spatially Modulated Vacua in Relativistic Field Theories,''
 arXiv:1706.02938 [hep-th].

\bibitem{Nitta:2017yuf-1} 
  M.~Nitta, S.~Sasaki and R.~Yokokura,
  ``Supersymmetry Breaking in Spatially Modulated Vacua,''
  Phys.\ Rev.\ D {\bf 96}, no. 10, 105022 (2017)
  [arXiv:1706.05232 [hep-th]].



    
\bibitem{Cecotti:1986gb} 
  S.~Cecotti and S.~Ferrara,
  ``Supersymmetric Born-infeld Lagrangians,''
  Phys.\ Lett.\ B {\bf 187}, 335 (1987).

\bibitem{Bagger:1996wp} 
  J.~Bagger and A.~Galperin,
  ``A New Goldstone multiplet for partially broken supersymmetry,''
  Phys.\ Rev.\ D {\bf 55}, 1091 (1997)
  [hep-th/9608177].

\bibitem{Rocek:1997hi} 
  M.~Rocek and A.~A.~Tseytlin,
  ``Partial breaking of global D = 4 supersymmetry, constrained superfields, and three-brane actions,''
  Phys.\ Rev.\ D {\bf 59}, 106001 (1999)
  [hep-th/9811232].



\bibitem{Kuzenko:2002vk-0} 
  S.~M.~Kuzenko and S.~A.~McCarthy,
  ``Nonlinear selfduality and supergravity,''
  JHEP {\bf 0302}, 038 (2003)
  [hep-th/0212039].
\bibitem{Kuzenko:2002vk-1} 
  S.~M.~Kuzenko and S.~A.~McCarthy,
  ``On the component structure of N=1 supersymmetric nonlinear electrodynamics,''
  JHEP {\bf 0505}, 012 (2005)
  [hep-th/0501172].

  \bibitem{Abe:2015nxa} 
  H.~Abe, Y.~Sakamura and Y.~Yamada,
  ``Matter coupled Dirac-Born-Infeld action in four-dimensional N=1 conformal supergravity,''
  Phys.\ Rev.\ D {\bf 92}, no. 2, 025017 (2015)
    [arXiv:1504.01221 [hep-th]].

\bibitem{Cecotti:1986jy} 
  S.~Cecotti, S.~Ferrara and L.~Girardello,
  ``Structure of the Scalar Potential in General $N=1$ Higher Derivative Supergravity in Four-dimensions,''
  Phys.\ Lett.\ B {\bf 187}, 321 (1987).

\bibitem{Farakos:2013zsa} 
F.~Farakos, S.~Ferrara, A.~Kehagias and M.~Porrati,
  ``Supersymmetry Breaking by Higher Dimension Operators,''
  Nucl.\ Phys.\ B {\bf 879}, 348 (2014)
  [arXiv:1309.1476 [hep-th]].
  
\bibitem{Kuzenko:2000tg-0} 
  S.~M.~Kuzenko and S.~Theisen,
  ``Supersymmetric duality rotations,''
  JHEP {\bf 0003}, 034 (2000)
  [hep-th/0001068].
\bibitem{Kuzenko:2000tg-1} 
  S.~M.~Kuzenko and S.~Theisen,
  ``Nonlinear selfduality and supersymmetry,''
  Fortsch.\ Phys.\  {\bf 49}, 273 (2001)
  [hep-th/0007231].

\bibitem{Kuzenko:2000tg-2} 
  S.~M.~Kuzenko,
  ``The Fayet-Iliopoulos term and nonlinear self-duality,''
  Phys.\ Rev.\ D {\bf 81}, 085036 (2010)
  [arXiv:0911.5190 [hep-th]].




\bibitem{Cecotti:1987sa} 
  S.~Cecotti,
  ``Higher Derivative Supergravity Is Equivalent To Standard Supergravity Coupled To Matter. 1.,''
  Phys.\ Lett.\ B {\bf 190}, 86 (1987).


\bibitem{Ferrara:2014fqa} 
  S.~Ferrara, A.~Kehagias and A.~Riotto,
  ``The Imaginary Starobinsky Model and Higher Curvature Corrections,''
  Fortsch.\ Phys.\  {\bf 63}, 2 (2015)
  [arXiv:1405.2353 [hep-th]].


\bibitem{Farakos:2017mwd} 
  F.~Farakos, S.~Ferrara, A.~Kehagias and D.~Lust,
  ``Non-linear Realizations and Higher Curvature Supergravity,''
  arXiv:1707.06991 [hep-th].


\bibitem{Diamandis:2017ems} 
  G.~A.~Diamandis, B.~C.~Georgalas, K.~Kaskavelis, A.~B.~Lahanas and G.~Pavlopoulos,
  ``Deforming the Starobinsky model in ghost-free higher derivative supergravities,''
  Phys.\ Rev.\ D {\bf 96}, no. 4, 044033 (2017)
  [arXiv:1704.07617 [hep-th]].


\bibitem{Freedman:2012zz} 
  D.~Z.~Freedman and A.~Van Proeyen,
  ``Supergravity'', Cambridge University Press, Cambridge
U.K., (2012).



\bibitem{Kallosh:2013xya} 
  R.~Kallosh and A.~Linde,
  ``Superconformal generalizations of the Starobinsky model,''
  JCAP {\bf 1306}, 028 (2013)
  [arXiv:1306.3214 [hep-th]].


\bibitem{Farakos:2013cqa} 
  F.~Farakos, A.~Kehagias and A.~Riotto,
  ``On the Starobinsky Model of Inflation from Supergravity,''
  Nucl.\ Phys.\ B {\bf 876}, 187 (2013)
  [arXiv:1307.1137 [hep-th]].

\bibitem{Ketov:2013dfa} 
  S.~V.~Ketov and T.~Terada,
  ``Old-minimal supergravity models of inflation,''
  JHEP {\bf 1312}, 040 (2013)
  [arXiv:1309.7494 [hep-th]].

\bibitem{Kaku:1978nz} 
  M.~Kaku, P.~K.~Townsend and P.~van Nieuwenhuizen,
  ``Properties of Conformal Supergravity,''
  Phys.\ Rev.\ D {\bf 17}, 3179 (1978).


\bibitem{Kaku:1978ea} 
  M.~Kaku and P.~K.~Townsend,
  ``Poincare Supergravity As Broken Superconformal Gravity,''
  Phys.\ Lett.\  {\bf 76B}, 54 (1978).


\bibitem{Townsend:1979ki} 
  P.~K.~Townsend and P.~van Nieuwenhuizen,
  ``Simplifications of Conformal Supergravity,''
  Phys.\ Rev.\ D {\bf 19}, 3166 (1979).


\bibitem{Kugo:1982cu} 
  T.~Kugo and S.~Uehara,
  ``Conformal and Poincare Tensor Calculi in $N=1$ Supergravity,''
  Nucl.\ Phys.\ B {\bf 226}, 49 (1983).

\bibitem{Ozkan:2014cua} 
  M.~Ozkan and Y.~Pang,
  ``$R^n$ Extension of Starobinsky Model in Old Minimal Supergravity,''
  Class.\ Quant.\ Grav.\  {\bf 31}, 205004 (2014)
  [arXiv:1402.5427 [hep-th]].

\bibitem{Ferrara:1983dh} 
  S.~Ferrara, L.~Girardello, T.~Kugo and A.~Van Proeyen,
  Nucl.\ Phys.\ B {\bf 223}, 191 (1983).

\bibitem{Cecotti:2014ipa} 
  S.~Cecotti and R.~Kallosh,
  ``Cosmological Attractor Models and Higher Curvature Supergravity,''
  JHEP {\bf 1405}, 114 (2014)
  [arXiv:1403.2932 [hep-th]].
\bibitem{Kallosh:2013yoa} 
  R.~Kallosh, A.~Linde and D.~Roest,
  JHEP {\bf 1311}, 198 (2013)
  doi:10.1007/JHEP11(2013)198
  [arXiv:1311.0472 [hep-th]].
\bibitem{Carrasco:2015uma} 
  J.~J.~M.~Carrasco, R.~Kallosh, A.~Linde and D.~Roest,
  Phys.\ Rev.\ D {\bf 92}, no. 4, 041301 (2015)
  doi:10.1103/PhysRevD.92.041301
  [arXiv:1504.05557 [hep-th]].
\bibitem{Sohnius:1981tp} 
  M.~F.~Sohnius and P.~C.~West,
  ``An Alternative Minimal Off-Shell Version of N=1 Supergravity,''
  Phys.\ Lett.\  {\bf 105B}, 353 (1981).

 \end{thebibliography}
\end{document}